\documentclass[a4paper,fleqn,usenatbib]{mnras}

\usepackage{booktabs,multirow}

\usepackage{amsmath}	
\usepackage{newtxtext,newtxmath}

\usepackage{caption}

\usepackage[T1]{fontenc}
\usepackage{ae}
\usepackage{aecompl}

\usepackage{graphicx,amssymb}
\usepackage{multicol}

\title[Environmental effects on SFGs in the COSMOS field]{Effect of the environment on star formation activity and stellar mass for star-forming galaxies in the COSMOS field} 

\author[S. M. Randriamampandry et al.]{S. M. Randriamampandry,$^{1,2}$\thanks{E-mail: solohery@saao.ac.za} M. Vaccari$^{3,4}$ and K. M. Hess $^{5,6}$  
\\
$^{1}$South African Astronomical Observatory,  P.O. Box 9, Observatory 7935, Cape Town, South Africa \\
$^{2}$A\&A, Department of Physics, Faculty of Sciences, University of Antananarivo, B.P. 906, Antananarivo 101, Madagascar \\
$^{3}$Department of Physics and Astronomy, University of the Western Cape, Robert Sobukwe Road, Bellville 7535, South Africa \\
$^{4}$INAF - Istituto di Radioastronomia, via Gobetti 101, I-40129 Bologna, Italy \\
$^{5}$ASTRON, the Netherlands Institute for Radio Astronomy, PO Box 2, 7990 AA Dwingeloo, The Netherlands \\
$^{6}$Kapteyn Astronomical Institute, University of Groningen, PO Box 800, 9700 AV Groningen, The Netherlands \\
}

\begin{document}

\date{Accepted 2020 August 26. Received 2020 August 23; in original form 2019 September 14}


\pagerange{\pageref{firstpage}--\pageref{lastpage}} \pubyear{2020}

\maketitle

\label{firstpage}

\begin{abstract}
We investigate the relationship between environment and the galaxy main sequence (the relationship between stellar mass and star formation rate) and also the relationship between environment and radio luminosity (P$_{\rm 1.4GHz}$) to shed new light on the effects of the environments on galaxies.
We use the VLA-COSMOS 3 GHz catalogue that consists of star-forming galaxies (SFGs) and quiescent galaxies (AGN) in three different environments (field, filament, cluster) and for three different galaxy types (satellite, central, isolated). We perform for the first time a comparative analysis of the distribution of SFGs with respect to the main sequence (MS) consensus region from the literature, taking into account galaxy environment and using radio observations at 0.1 $\leq$ z $\leq$ 1.2. 
Our results corroborate that SFR is declining with cosmic time which is consistent with the literature. We find that the slope of the MS for different $z$ and M$_{*}$ bins is shallower than the MS consensus with a gradual evolution towards higher redshift bins, irrespective of environments.  We see no SFR trends on both environments and galaxy type given the large errors. In addition, we note that the environment does not seem to be the cause of the flattening of MS at high stellar masses for our sample.

\end{abstract}


\section{Introduction} \label{Introduction}
Star formation (SF) is one of the principle indicators of the evolution of galaxies. SF can be used as one of the main tracer of galaxy evolution as it allows us to characterize the newly born stars in the galaxy by measuring the star formation rate (SFR)  i.e. the amount of the total mass of stars formed per year. Furthermore, star formation history (SFH) is important as it enables us to find changes in the SFR across cosmic time hence traces the evolution of the galaxy. An estimate of the integrated SFH of a galaxy can be inferred from stellar mass \citep[see][]{2000ApJ...536L..77B,2004Natur.428..625H,2005ApJ...622L...5T} via the relationship between star formation and the growth of stellar mass; defined as specific SFR  (sSFR = SFR/M$_{*}$ i.e. the star formation rate per unit stellar mass).  

Star formation activity further correlates strongly with galaxy stellar mass (M$_{*}$): a relationship that was coined for the first time by \cite{2007ApJ...660L..43N} -- as the star formation main sequence (MS), and has been extensively studied since \cite[see][]{2007ApJ...670..156D,2009ApJ...698L.116P,2010ApJ...714.1740M,2011ApJ...739L..40R,2012ApJ...754L..29W,2013ApJ...763..129S,2014ApJ...791L..25S,2015A&A...579A...2I,2017ApJ...840...47B,2018A&A...615A.146P,2019MNRAS.483.3213P}. Star-forming galaxies (SFGs) along the MS relation have star formation self-regulated by secular processes and at a level dictated predominantly by their stellar masses \cite[e.g.][]{2014ApJ...795..104W,2015ApJ...798..115L}. The MS relation implies that star formation rate histories over an averaged population of SFGs are mostly regular, smooth and deceasing with M$_{*}$-dependent timescale \cite[e.g.][]{2004Natur.428..625H} prior to the shutdown of star formation. 

The universality of the observed MS relation for SFGs has been investigated for over a decade and found to hold for a wide range of redshift from the local universe at z $\simeq$ 0 \cite[see][]{2004MNRAS.351.1151B,2007ApJS..173..267S}, to z $\simeq$ 1 \cite[e.g.][]{2007ApJ...660L..43N, 2007A&A...468...33E,2015A&A...579A...2I}, z $\simeq$ 2 \cite[see][]{2007ApJ...670..156D,2009ApJ...698L.116P,2011ApJ...739L..40R,2012ApJ...754L..29W,2015A&A...580A..42K}, and from z $\simeq$ 3 \cite[e.g.][]{2010ApJ...714.1740M,2017ApJ...840...47B} to z $\simeq$ 4  \cite[e.g.][]{2009ApJ...694.1517D,2015ApJ...807..141P}, and even up to higher z $\sim$ 7 \cite[see][]{2012ApJ...754...83B,2013ApJ...763..129S, 2014ApJ...791L..25S, 2015ApJ...799..183S}. Furthermore, there are studies of MS that have already been conducted at 0.2 $\leq$ z < 6 with FIR Herschel SPIRE \cite[e.g.][]{2018A&A...615A.146P} and hence the relationship has been in place for at least 90\% of the age of the Universe. Therefore, it can be used to characterise how the instantaneous star formation is determined by the past star formation histories of an individual or a subset of population of SFGs.

However, other studies suggest that the SFR in SFGs can be regulated and quenched by different physical processes linked to environments such as galaxy interactions, and galaxy minor/major mergers \cite[e.g.][]{2010ApJ...721..193P,2013ApJ...778...23G,2020ApJ...889..156C} that may alter the properties of the MS relation. Despite of these processes, and despite different methods used to measure the galaxy properties, the MS relation still holds. Furthermore, its characteristics have been reproduced in numerical simulations up to at least z$\sim$3 \cite[e.g.][]{2006ApJ...639..672F,2010MNRAS.404.1355D,2011MNRAS.415...11D,2016MNRAS.457.2790T}. Previous studies that have looked at the MS of SFGs in different environments found that cluster galaxies and those found in groups have lower sSFR compared to field galaxies \citep[e.g.][]{2010ApJ...710L...1V,2013ApJ...775..126H,2014ApJ...782...33L,2017ApJ...845...74J}. The cluster, the field, and the filament environments of galaxies shape their properties such as colour and mass thus playing important roles in their evolution over cosmic time \citep[e.g.][]{2006MNRAS.373..469B,2012ApJ...746..188M,2018PASJ...70S..23J} and hence in their ultimate fate.

In this work we targeted a large population of SFGs at 0.1 < z < 1.2 from one of the deepest radio surveys publicly available with a "clean" separation of SFGs from AGN. We used 1.4 GHz luminosity which traces the synchrotron emission from SFGs. The non-thermal radio luminosity can be an excellent dust-free estimator of star formation \cite[see][]{1992ARA&A..30..575C,2003ApJ...586..794B,2011ApJ...737...67M} and further here we attempt to use it to look at the behavior of the radio luminosities of our larger radio-selected sample of SFGs across different environments. Investigation the MS relation as well as the radio luminosity of SFGs as a function of the environment and galaxy type may illuminate the role of environmental processes in galaxy evolution. 
This work aims to study the relationship between star formation rate (SFR) and stellar mass (M$_{*}$) of star-forming galaxies, and also the relationship between environment and radio luminosity (P$_{\rm 1.4GHz}$), to shed new light on the effects of the environments (field, filament, cluster) on various galaxy types (satellites, central, isolated) at 0.1 $\leq$ z $\leq$ 1.2. To the best of our knowledge, other no work has looked at the MS taking advantage of one of the deepest available radio surveys. Considering multi-wavelength data for these galaxies and the available measurements of their environments has allowed us to investigate the effect of the environment on the galaxy main sequence in a unique way.

The paper is organised as follows. Section \ref{Sample and Galaxy Properties} presents our sample of SFGs and their properties. Section \ref{Results} presents our results. Finally, Section \ref{Discussion} and Section \ref{Conclusions} discuss and summarise our findings, and  suggest some future work. Throughout this paper, we adopt $\rm H_{0}$ = 71 km s$^{-1}$ Mpc$^{-1}$, $\Omega_{\rm m}$ = 0.27 and $\Omega_{\rm DE}$ = 0.73.

\section{Sample and Galaxy Properties} \label{Sample and Galaxy Properties}
For our analysis, we take advantage of data from the Cosmic Evolution Survey \cite[COSMOS\footnote{\url{http://cosmos.astro.caltech.edu/page/astronomers}},][]{2007ApJS..172....1S} field in the available redshift range of 0.1 $\leq$ z $\leq$ 1.2. The COSMOS is a multi-wavelength survey that covers two square degrees field of view, and is centered at RA = +150.119167 and DEC = +2.205833. The survey aims to study galaxy formation and evolution as a function of redshift, out to z $\sim$ 5, and the large scale structure environment. The details of COSMOS survey and observations are described in \cite{2007ApJS..172....1S}. In the next sections, we provide a description of our sample and its properties, and also indicate where they particularly come from.

\subsection{Sample}
In this work, we used a sample of radio selected star-forming galaxies drawn from the multiwavelength catalogue\footnote{\url{http://jvla-cosmos.phy.hr/dr1/}} compiled by \cite{2017A&A...602A...2S} which is based on the VLA-COSMOS 3 GHz Large Project radio source catalogue \citep{2017A26A...602A...1S}. We complemented the \cite{2017A&A...602A...2S} catalogue with two ancillary catalogues that consist of the cosmic web environment\footnote{\url{https://irsa.ipac.caltech.edu/data/COSMOS/tables/environment/}} catalogue of \cite{2015ApJ...805..121D,2017ApJ...837...16D} and a photometric catalogue from the COSMOS survey conducted in 2015  \cite[COSMOS2015\footnote{\url{ftp://ftp.iap.fr/pub/from_users/hjmcc/COSMOS2015/}},][]{2016ApJS..224...24L} to form our sample. 

We matched the \cite{2017A&A...602A...2S} catalogue to two further catalogues. The first of this is the cosmic web environment catalogue of \cite{2015ApJ...805..121D,2017ApJ...837...16D} in aiming to study the effects of the environment (field, filament, cluster) on star formation main sequence for various type of galaxies (satellite, central, isolated). Secondly we matched to the COSMOS2015 to obtain stellar mass (M$_{*}$) and photometric redshifts ($z$) for each galaxy. We did all the matching using COSMOS2015 source identification numbers.

\begin{figure}
\centering
\includegraphics[width=.45\textwidth]{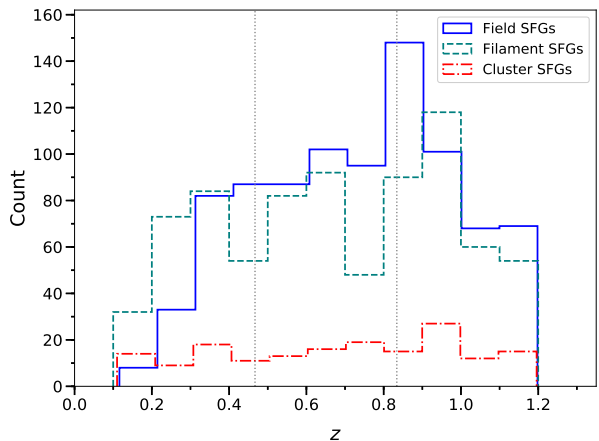} 
 \caption{Histogram of the redshifts of our sample. The three different environment estimators from \protect\cite{2017ApJ...837...16D} are shown in blue solid line for field, in cyan dashed line for filament, and in red dashed-dotted line for cluster SFGs. The grey dotted lines indicate the bin width of each sub-sample that we work throughout the paper. (A colour version of this figure is available in the online journal.)}
 \label{fig: redshift}
 \end{figure}

The final number of sources out of these matching procedures resulted in 2568 radio detections with counterparts in the other catalogues (note that we eliminate 'bad data' that has FLAG = 1 in the COSMOS2015 catalogue) where 1836 are SFGs and 732 are AGN.  \cite{2017A&A...602A...2S} used X-ray, MIR, and Spectral Energy Distribution (SED)-based observations to separate AGN/SFGs plus a combined rest frame colour and radio excess diagnostic to obtain a "clean" sample of SFGs. We removed those AGN from our sample and focus on only the SFGs throughout the rest of this paper.

Table \ref{tab: env-pos} summarizes the number of galaxies by their respective environments and galaxy types. Figure \ref{fig: redshift} shows the redshift distribution of our sample where the three different environment estimators from \cite{2017ApJ...837...16D} are indicated with a blue solid line for field, cyan dashed line for filament, and red dashed-dotted line for cluster SFGs.   

\begin{table}
\caption{The breakdown of the total number of galaxies for each cosmic web environments (field, filament, cluster) and for each galaxy types (satellite, central, isolated) in our sample.}
\begin{center} 
\begin{tabular}{c c c} \hline \hline
 Field   & Filament   & Cluster       \\   \hline 
 880   &  787          &  169        \\ \hline \hline 
 Satellite        &     Central            &        Isolated        \\ \hline 
 697        &     450          &        689       \\ \hline
\end{tabular}
\end{center}
\label{tab: env-pos}
\end{table}

\subsection{Properties of galaxies} 
\begin{figure}
\centering
\includegraphics[width=.45\textwidth]{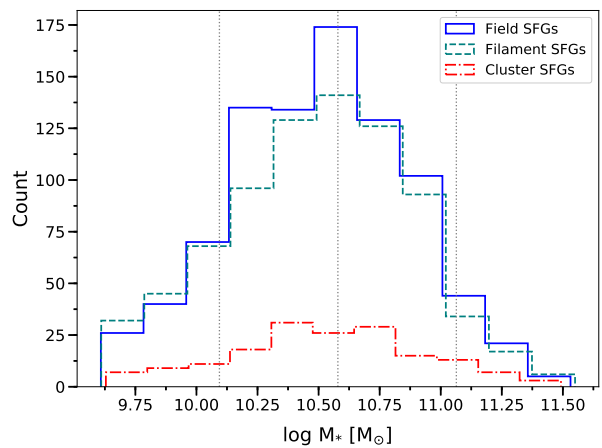} 
 \caption{Histogram of stellar masses of our sample. The three different environment estimators from \protect\cite{2017ApJ...837...16D} are shown in blue solid line for field, in cyan dashed line for filament, and in red dashed-dotted line for cluster SFGs.  The grey dotted lines indicate the bin width of each sub-sample that we work throughout the paper.(A colour version of this figure is available in the online journal.)}
 \label{fig: Mstar}
 \end{figure}

We use the derived M$_{*}$ from the COSMOS2015 catalogue  which are based on SED-fitting using the Stellar Population Synthesis (SPS) model of \cite{2003MNRAS.344.1000B} (BC03) templates by assuming an initial mass function (IMF) of \cite{2003PASP..115..763C}.   The full details of the method for estimating the M$_{*}$ are presented in \cite{2015A&A...579A...2I,2016ApJS..224...24L}. We show the histogram of the M$_{*}$  in Figure \ref{fig: Mstar} where the three different environment estimators from \cite{2017ApJ...837...16D} are indicated in blue solid line for field, in cyan dashed line for filament, and in red dashed-dotted line for cluster SFGs. We  note that the M$_{*}$ normalized distributions are similar across the different environments. 

\begin{figure}
\centering
\includegraphics[width=.45\textwidth]{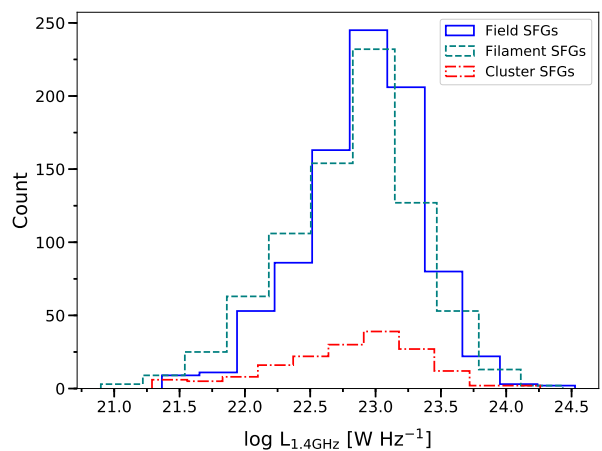} 
 \caption{Histogram of radio luminosity of our sample. The three different environment estimators from \protect\cite{2017ApJ...837...16D} are shown in blue solid line for field, in cyan dashed line for filament, and in red dashed-dotted line for cluster SFGs. (A colour version of this figure is available in the online journal.)}
  \label{fig: Lradio}
 \end{figure}

The SFRs were measured by \cite{2017A&A...602A...2S} which were estimated from the total infrared luminosity by using the \cite{1998ApJ...498..541K} conversion factor that were scaled to a \cite{2003PASP..115..763C} IMF. In addition, the rest-frame 1.4 GHz radio luminosity were measured from the 1.4--3 GHz spectral index, when available, or assuming a spectral index\footnote{$\alpha$ is the power law slope of the synchrotron radiation, and is defined as S$_{\upnu}$$\sim$$\upnu ^{- \alpha}$ where S$_{\upnu}$ and $\upnu$ are the flux density and frequency, respectively.} of $\alpha = 0.7$.  Figure \ref{fig: Lradio} shows the distribution of 1.4 GHz radio luminosity (P$_{\rm 1.4GHz}$) of our sample. The full details of the method for measuring the total infrared luminosity, SFRs, and P$_{\rm 1.4GHz}$ are presented in \cite{2017A&A...602A...2S}.

Finally, the measurements of local environments are based on density field that are constructed via weighted adaptive kernel smoothing estimator \citep{2015ApJ...805..121D,2017ApJ...837...16D}. The components of the cosmic web environments (filament, cluster, and field) are then extracted from the constructed density field through the Hessian matrix technique  \cite[see][]{2017ApJ...837...16D}. While the classification of galaxy types (central, satellite, and isolated) are observationally identified through galaxy groups where the most massive galaxy in each single group selected as a central and the rest as satellites or otherwise isolated if not associated with any groups.  Groups of galaxies are identified using the commonly used friends-of-friends algorithm \citep{1982ApJ...257..423H,2014MNRAS.440.1763D,2017ApJ...837...16D}. 
We refer the readers to the work of \cite{2015ApJ...805..121D,2017ApJ...837...16D} for full details on the cosmic web measurements and the classification of galaxy types.

\section{Results} \label{Results}

\subsection{MS of SFGs and environments}

\subsubsection{Aim and Approach}
Our primary aim is to look at the differences of the MS of SFGs with respect to environment as a function of redshift, and to do so, we will be comparing our results to the MS consensus of \cite{2014ApJS..214...15S}. \cite{2014ApJS..214...15S} have collated and calibrated 25 MS studies of SFGs that are taken from a range of multiwavelength observations (UV-to-radio) from the literature to generate a MS consensus. The MS consensus at $\pm$0.2 dex dispersion is compiled based on the M$_{*}$ and SFR measured using BC03 SPS model with Kroupa IMF \citep{2014ApJS..214...15S}. We choose their MS consensus as a benchmark to which we compare our work as the galaxy properties of our sample were measured with the same SPS model. The chosen MS consensus enables us to minimise some systematics due to sample selection effects, SFR indicators, and methods used to locate the MS when comparing different results from the literature. 

As our results such as M$_{*}$ were measured via SED-fitting through $\chi^{2}$ method based on best-fit template using optical-NIR photometry and BC03 SPS model specifically using a Chabrier IMF. For conformity of comparisons, we convert M$_{*}$, SFR and the MS consensus and any others MS from the literature into Chabrier IMF. We use a conversion of IMF for both M$_{*}$ and SFR by following the equation (2) of \cite{2014ApJS..214...15S} which is defined as $\rm Kroupa = 1.06 \times Chabrier = 0.62 \times Salpeter$. We adapt this relation for the M$_{*}$ and SFR by assuming that the M$_{*}$ and SFR have similar relative offsets. A similar ratio has been observed for the SFR via a conversion from Salpeter SFR \citep{1998ARA&A..36..189K} to Kroupa SFR \citep{2012ARA&A..50..531K}.  We also use SFR in this work instead of sSFR for direct comparison to the control sample from \cite{2014ApJS..214...15S}. Throughout the paper, we take into account this conversion when comparing M$_{*}$ and SFR from any other studies.

To disentangle any redshift and or M$_{*}$ dependence on the MS for a given environment, we have divided our sample of SFGs into three bins of equal width such that the first bin is at 0.10 $\leq$ z < 0.47 and the second bin is at 0.47 $\leq$ z < 0.83 and the third bin is at 0.83 $\leq$ z $\leq$ 1.20. Likewise, we split sources into four M$_{*}$ bins with a width of approximately 0.5 dex such that the first bin is for 9.6 $\leq$ log M$_{*}$ < 10.1 and the second bin is for 10.1 $\leq$ log M$_{*}$ < 10.6, and the third bin is for 10.6 $\leq$ log M$_{*}$ < 11.1 and the last bin is for 11.1 $\leq$ log M$_{*}$ <  11.6. 

\begin{table}
\caption{The breakdown of the total number of galaxies for each $z$ and M$_{*}$ bins.}
\begin{center} 
\begin{tabular}{l c } \hline \hline
$z$ bins  & No of sources        \\   \hline 
1$^{\rm st}$ (0.10 $\leq$ z < 0.47)   &   435          \\ \hline 
2$^{\rm nd}$ (0.47 $\leq$ z < 0.83)    &    697        \\ \hline 
3$^{\rm rd}$ (0.83 $\leq$ z $\leq$ 1.20)   &   704    \\ \hline \hline 
M$_{*}$ bins      & No of sources         \\ \hline 
1$^{\rm st}$ (9.6 $\leq$ log M$_{*}$ < 10.1)   &   275   \\ \hline 
2$^{\rm nd}$ (10.1 $\leq$ log M$_{*}$ < 10.6)    &    812   \\ \hline 
3$^{\rm rd}$ (10.6 $\leq$ log M$_{*}$ < 11.1)   &   647   \\ \hline \hline 
4$^{\rm th}$ (11.1 $\leq$ log M$_{*}$ < 11.6)   &   102   \\ \hline \hline 

\end{tabular}
\end{center}
\label{tab: bins}
\end{table}

Throughout the work, we refer to these bins as the lower, intermediates, and the higher redshift or M$_{*}$ bins, respectively.  
Table \ref{tab: bins} presents the total number of galaxies for each $z$ and M$_{*}$ bins.  We note that although these bins have similar widths, when interpreting our results we strive to take into account the number of sources in each bin to minimise bias. 

To disentangle completeness limit, in a similar way to the method of \citet[see Figure 5]{2015MNRAS.453.1079B}, we use the empirical relationship between radio luminosity and SFR of \cite{2017A&A...602A...4D} to estimate the SFR limits. Using the minimum P$_{\rm 1.4GHz}$ in each single redshift bin, SFR limit is defined as: 

\begin{equation*}
\rm SFR \ [M_{\odot}\ yr^{-1}] = f_{IMF} 10^{-24} 10^{q_{TIR}(z,\alpha)} \times P_{1.4GHz} \ [W\ Hz^{-1}],
\end{equation*}
\rm where

\begin{equation*}
q_{\rm TIR}(z) = \begin{cases}
(2.88 \pm 0.03)(1 + z)^{-0.19 \pm 0.01} & \rm for \ \alpha = 0.7 \\
(2.85 \pm 0.03)(1 + z)^{-0.22 \pm 0.01}  & \rm for \ \alpha = 0.8
\end{cases}
\end{equation*}

where f$_{\rm IMF}$ is a factor accounting for the assumed initial mass function (f$_{\rm IMF}$ = 1 for a Chabrier IMF). The q$_{\rm TIR}$(z,$\alpha$) is the infrared-to-1.4 GHz radio luminosity ratio where z and $\alpha$ are the average redshift in each bin and the average assumed spectral index of the SFGs population, respectively.

\begin{figure*}
\begin{multicols}{2}
    \includegraphics[width=0.97\linewidth]{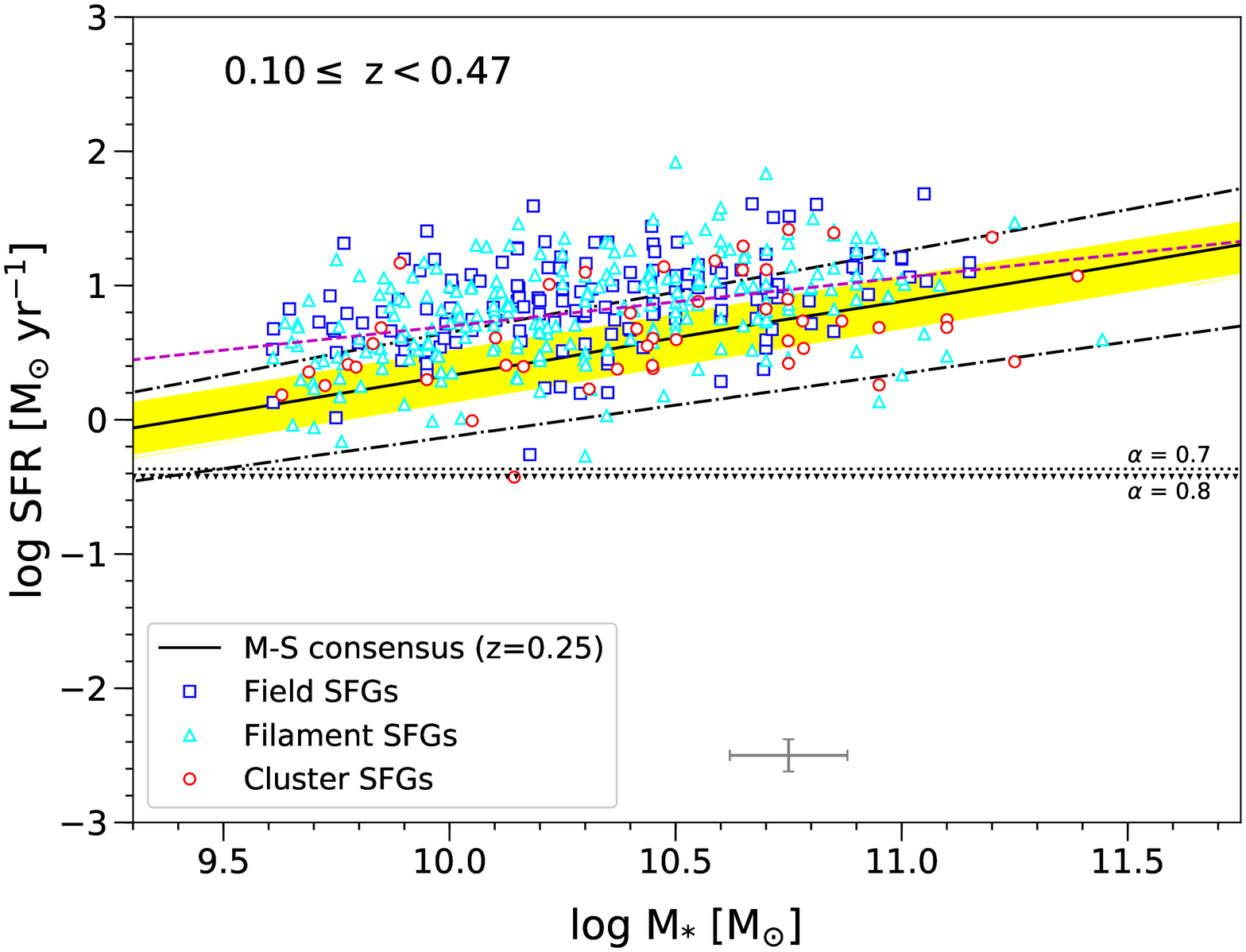}\par 
    \includegraphics[width=0.97\linewidth]{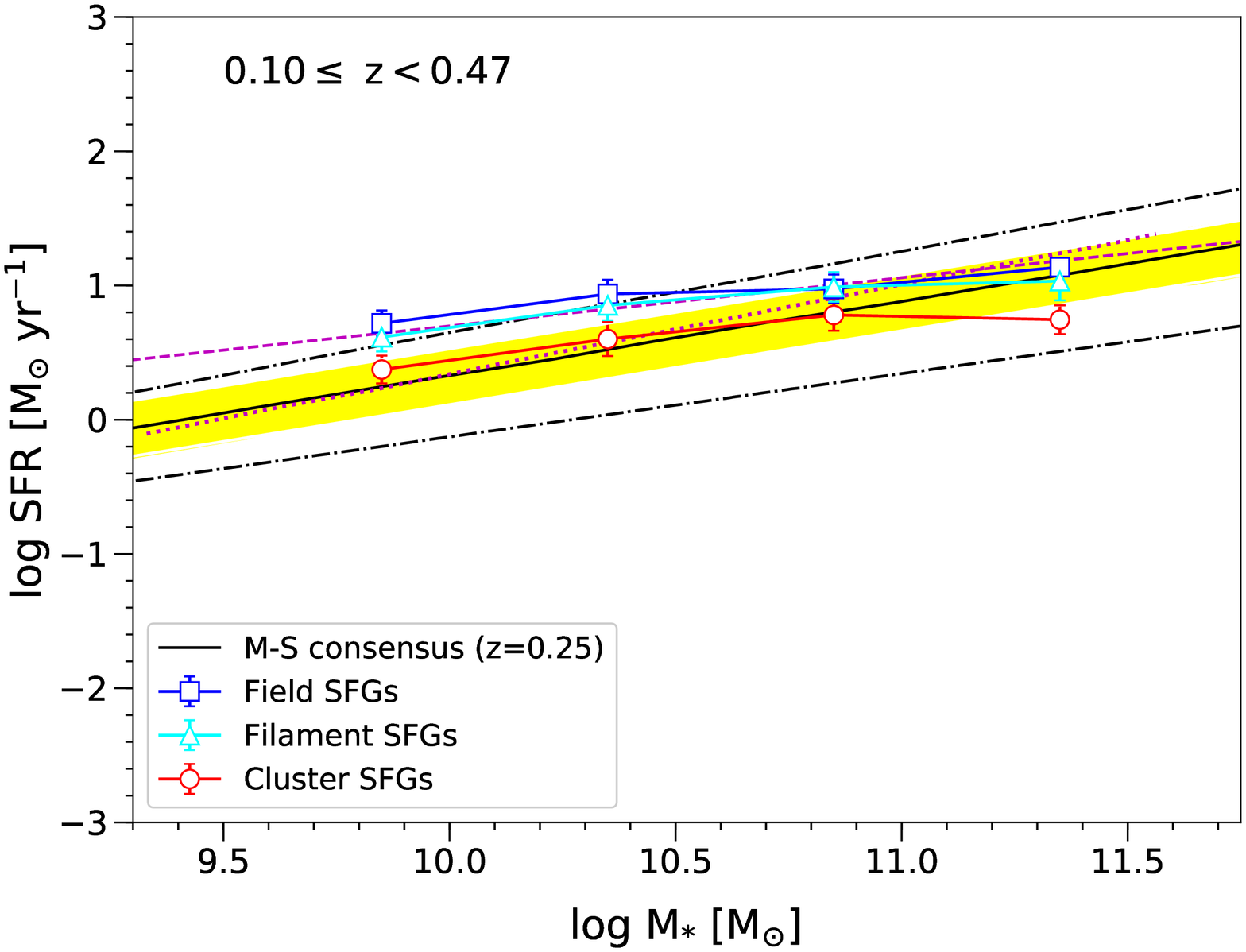}\par
\end{multicols}
\begin{multicols}{2}
    \includegraphics[width=0.97\linewidth]{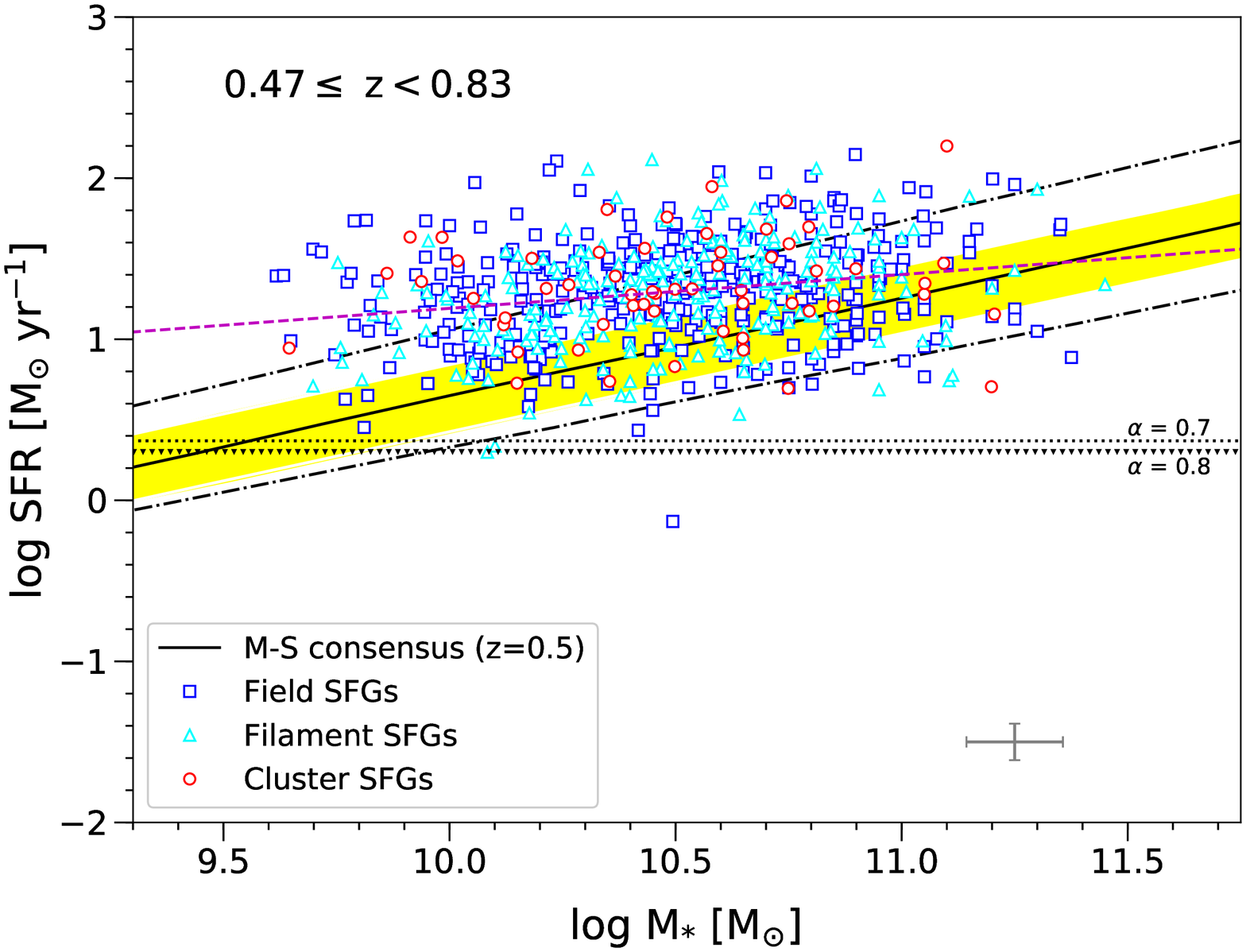}\par
    \includegraphics[width=0.97\linewidth]{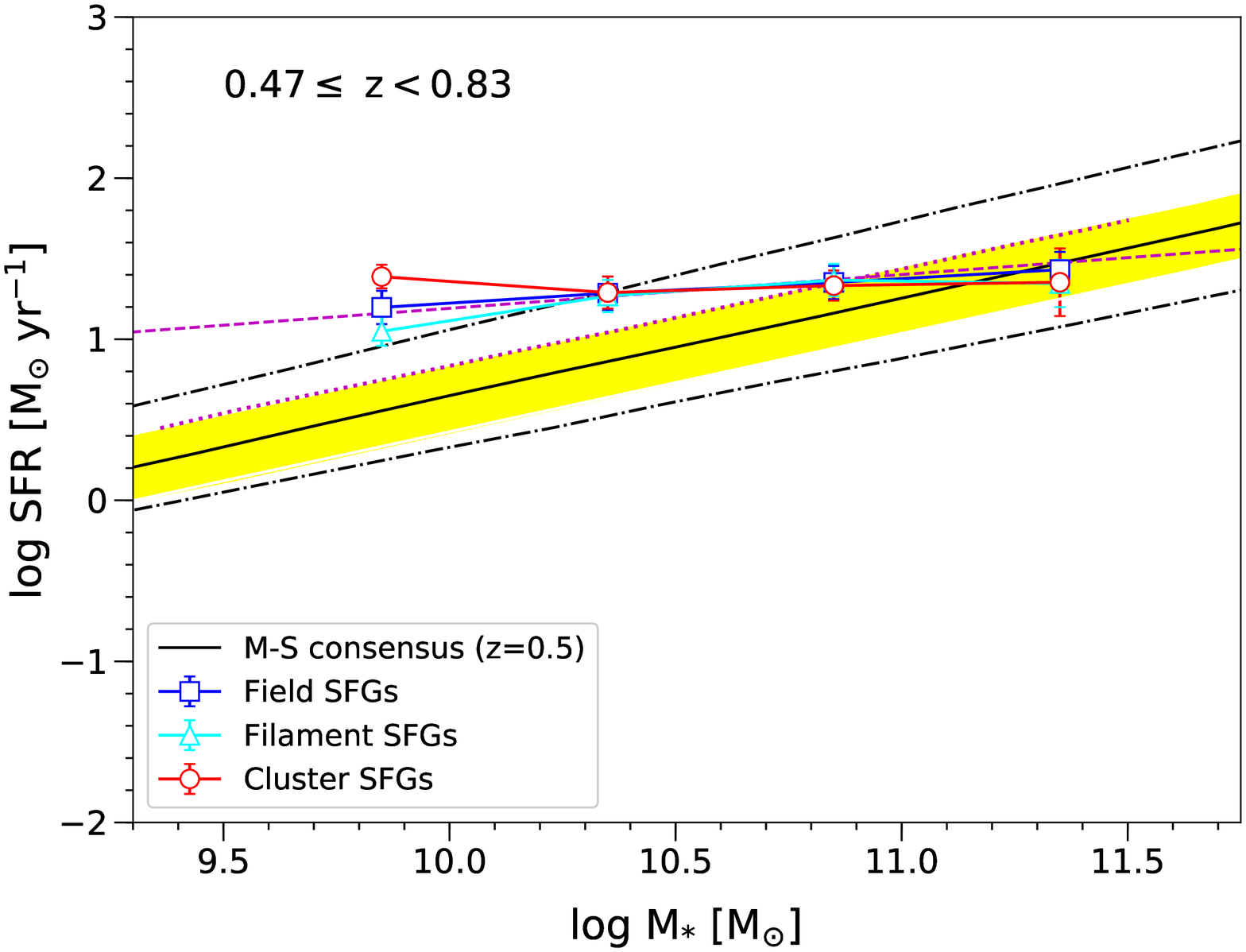}\par
\end{multicols}
\begin{multicols}{2}
    \includegraphics[width=0.97\linewidth]{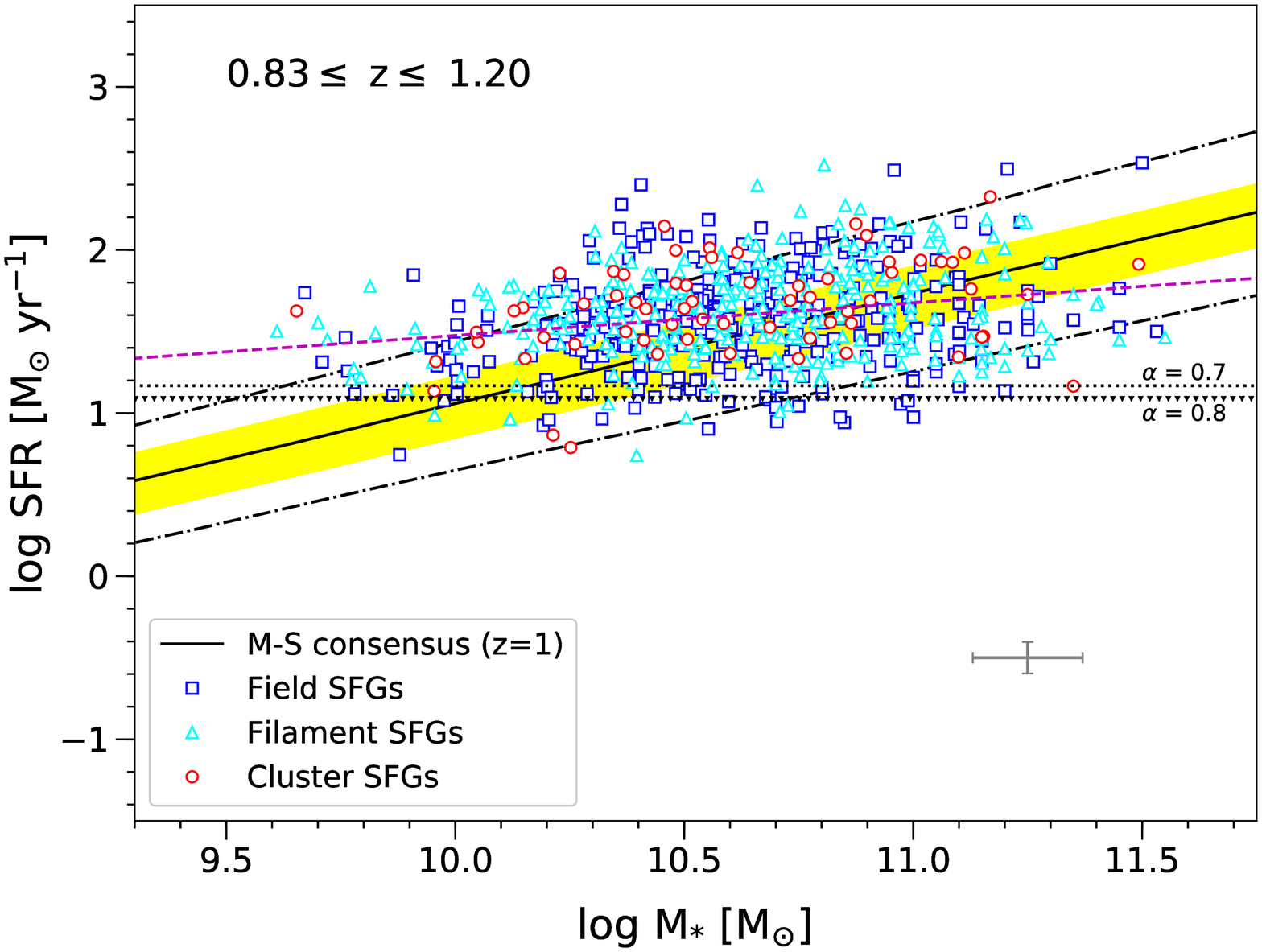}\par
    \includegraphics[width=0.97\linewidth]{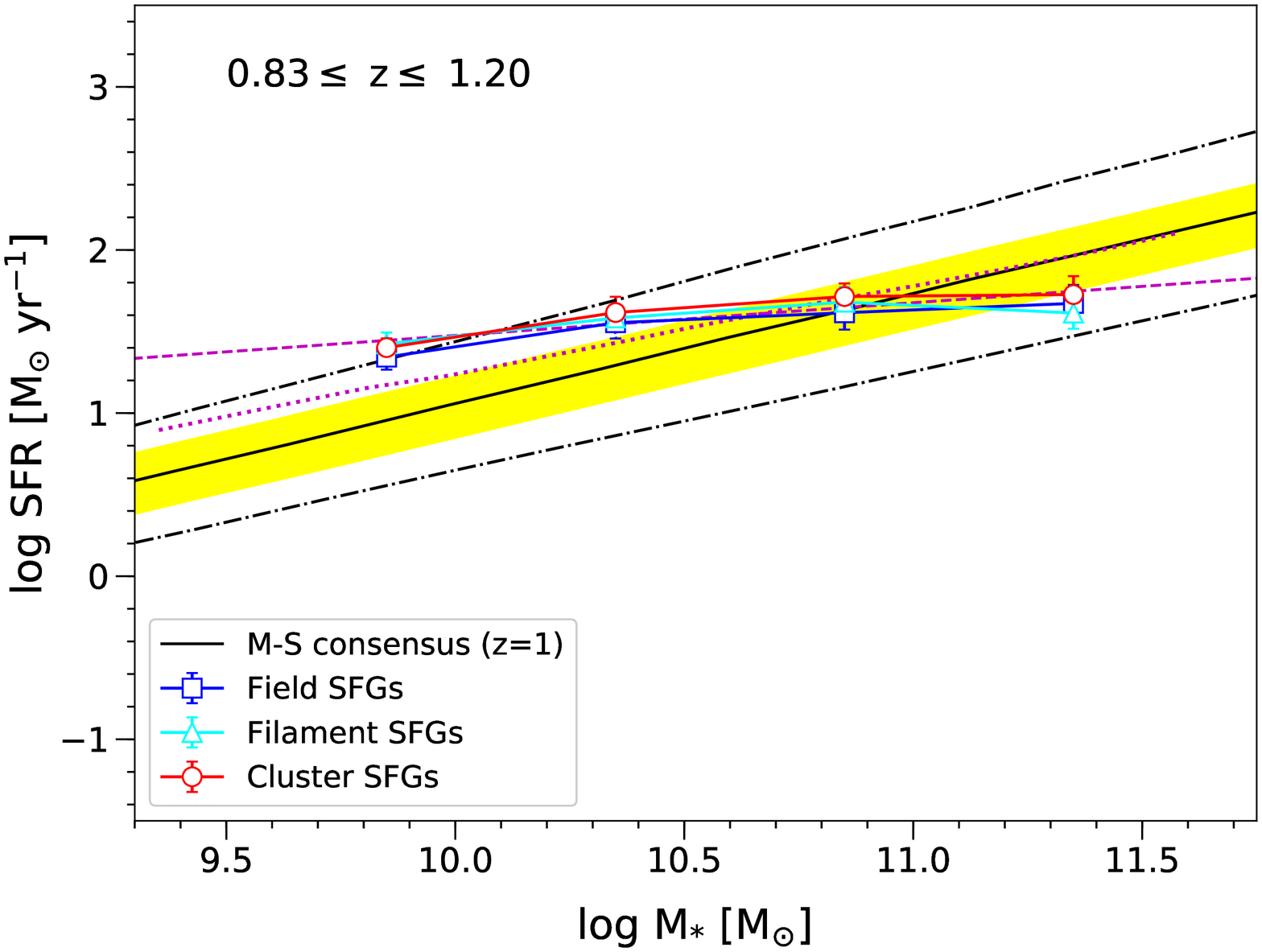}\par
\end{multicols}
\caption{Star formation rate (SFR) against stellar mass (M$_{*}$) of SFGs for the three different environments for the lower z ({\it top panel}), intermediate ({\it middle panel}), and higher ({\it bottom panel}) z bins. In each panel the blue square, teal triangle, and red circle symbols represent field, filament, and cluster SFGs, respectively. The black solid line of each panel indicates the MS consensus of \protect\cite{2014ApJS..214...15S} at z=0.25 ({\it top}), z=0.5 ({\it middle}), and z=1 ({\it bottom}) while the shaded yellow region indicates a scatter of $\pm$0.2 dex consensus dispersion. The black dashed-dotted lines indicate MS consensus of \protect\cite{2014ApJS..214...15S} at z=0 and 0.5 ({\it top}), z=0.25 and 1 ({\it middle}), and z=0.5 and 2 ({\it bottom}), respectively, which are shown as benchmark for comparison. The magenta dashed line shows the best-fit to the entire population of source (irrespective of environment). The ({\it left-hand}) panel shows the scatter plot of the SFGs upon the MS consensus.  The black dotted (for $\alpha$=0.7) and triangle down (for $\alpha$=0.8) lines indicate the estimated SFR limits calculated for each redshift bin and it is based on SFR and radio luminosity model of \protect\cite{2017A&A...602A...4D}.  The grey solid crossed-lines ({\it Left hand panel, bottom right}) represent the error bars which correspond to average 1$\upsigma$ errors based on the standard error of the mean. The ({\it right-hand}) panel presents the behaviour of the average M$_{*}$ of the four M$_{*}$ bins for the three environments. For further comparison, the MS best-fit from \protect\cite{2012ApJ...754L..29W} for each similar z bin is also shown in magenta dotted line. (A colour version of this figure is available in the online journal.)}
\label{fig: all_sf-ms}
\end{figure*}

The relationship between the star formation rate and the stellar mass of star forming galaxies for the three different environments is shown in Figure \ref{fig: all_sf-ms}.  The black solid line in each panel indicates the best-fit MS consensus of \cite{2014ApJS..214...15S} at $z=0.25$ ({\it top}), $z=0.5$ ({\it middle}), and $z=1$ ({\it bottom}) while the shaded yellow region indicates a scatter of $\pm$0.2 dex consensus dispersion. The black dashed-dotted lines indicate MS consensus of \cite{2014ApJS..214...15S} at $z=0$ and 0.5 ({\it top}), $z=0.25$ and 1 ({\it middle}), and $z=0.5$ and 2 ({\it bottom}), respectively, which are shown as benchmark for comparison.

In each panel of Figure \ref{fig: all_sf-ms} ({\it left-hand}), the black dotted (and triangle down) horizontal lines indicate the estimated SFR limits. In each panel of Figure \ref{fig: all_sf-ms} ({\it right-hand}), we plotted the mean of the SFR (which agrees with the median within $\approx$ 0.2 dex) and the errors shown are based on the standard error of the mean. We additionally compare our results to the MS best fitting lines from \cite{2012ApJ...754L..29W} in order to further quantify the slopes of our data as also shown in Figure \ref{fig: all_sf-ms} ({\it right-hand}). For conformity of comparisons, we choose to further compare to the MS from \cite{2012ApJ...754L..29W} as it is a similar  to the MS consensus of \cite{2014ApJS..214...15S}.

\subsubsection{Results of MS of SFGs}
In Figure \ref{fig: all_sf-ms}, we plot the MS of our binned sample where each row represents a different redshift bin. The left column presents the individual data points coloured by galaxy environments and the right column presents the mean trends of each environment, binned by galaxy stellar mass. The slopes and intercepts from our best fit lines are shown in Table \ref{tab: slopes}.

\begin{table}
\caption{A table summarizing the calculated slopes and intercepts for each $z$ bins.}
\begin{center} 
\begin{tabular}{l c c } \hline \hline
$z$ bins  & Slopes  & Intercepts        \\   \hline 
1$^{\rm st}$ (0.10 $\leq$ z < 0.47)  &   0.360$\pm$0.042   & 2.899$\pm$0.431          \\ \hline 
2$^{\rm nd}$ (0.47 $\leq$ z < 0.83)     & 0.210$\pm$0.034   &   0.905$\pm$0.359       \\ \hline 
3$^{\rm rd}$ (0.83 $\leq$ z $\leq$ 1.20)  &  0.201$\pm$0.031  &  0.530$\pm$0.325      \\ \hline 
\end{tabular}
\end{center}
\label{tab: slopes}
\end{table}

The top panel of Figure \ref{fig: all_sf-ms} ({\it left-hand}) shows that most of our sources are above the MS consensus in this lower redshift bin (0.10 $\leq$ z < 0.47) regardless of the environments. These galaxies located at the upper envelope of the MS consensus are actively forming stars i.e. sources that are further up from the yellow region. They seem to be observed to fit better to the benchmark MS consensus line at z = 0.5 across the M$_{*}$ range. In addition, we find that the slope of our best-fit shown in dashed line, irrespective of environments at 0.10 $\leq$ z < 0.47, seems to be shallower than the MS consensus. 

We also see that one of the cluster SFGs and one field SFGs at 0.47 $\leq$ z < 0.83 ({\it middle panel, left-hand}) are below the estimated SFR limits shown in the dotted horizontal lines. Furthermore, we see that quite a few number of the sources are below the estimated SFR limit at 0.83 $\leq$ z $\leq$ 1.20 ({\it bottom panel, left-hand}). For these radio sources detected under SFR limits, regardless of their environments, observed at the highest redshift bin appear to be associated with the spectral index as the two lines seem to depend on the assumed spectral index (i.e. $\alpha$ = 0.7 and $\alpha$ = 0.8) as well as cosmic time. The sensitivity limits deepen as the spectral index steepen when redshift increases. See also Section \ref{Possible caveats} for further discussion on SFR limits.

The top panel of Figure \ref{fig: all_sf-ms}  ({\it right-hand}) examines the behaviour of the MS as a function of the average M$_{*}$ for four M$_{*}$ bins. The slope of the MS for each respective environment appears to be shallower when compared to both the MS consensus and the best fit of \cite{2012ApJ...754L..29W} at similar redshift bin (0 < z < 0.5). For cluster SFGs, regardless of the flat slope towards the highest z bin, we may find a better agreement between our MS against the MS consensus. Nevertheless, both the field and filament SFGs do not agree with the yellow region at lower M$_{*}$ bins (9.6 $\leq$ log M$_{*}$ < 10.1 and 10.1 $\leq$ log M$_{*}$ < 10.6).

In the middle panel of Figure \ref{fig: all_sf-ms}, we show the MS for the intermediate redshift bin (0.47 $\leq$ z < 0.83). The middle panel of Figure \ref{fig: all_sf-ms} ({\it left-hand}) shows that there are still sources above the MS consensus regardless of the environments where most sources are up above the consensus dispersion yellow region across the M$_{*}$ range. We observe that the slope of our best-fit shown in dashed line may also be even shallower than the MS consensus in this bin at 0.47 $\leq$ z < 0.83. 

In the middle panel of Figure \ref{fig: all_sf-ms} ({\it right-hand}), we notice that the average SFR for the lowest bin is all above the MS consensus benchmark line and could be due to averaging lower number of sources in this bin. The slope of the MS for each respective environment may also be even shallower when compared to both the MS consensus and the best fit of \cite{2012ApJ...754L..29W} at similar redshift bin (0.5 < z < 1). The best fit line from \cite{2012ApJ...754L..29W} and the MS consensus line have the same slope and are in agreement within the consensus dispersion. We find that the mean SFR for all sources do not agree with the yellow region at lower M$_{*}$ bins (9.6 $\leq$ log M$_{*}$ < 10.1 and 10.1 $\leq$ log M$_{*}$ < 10.6) as well. 

At the bottom panel of Figure \ref{fig: all_sf-ms}, we present the MS for the highest redshift bin (0.83 $\leq$ z $\leq$ 1.20). The bottom panel of Figure \ref{fig: all_sf-ms} ({\it left-hand}) shows that regardless of the environments our sources appear to be clustered around the MS consensus and a higher fraction of them seem to be located in the MS consensus shaded yellow region. We find that the slope of our best-fit shown in dashed line may even be more shallow than the MS consensus. The bottom panel of Figure \ref{fig: all_sf-ms} ({\it right-hand}), the slope of the MS for each respective environment may also be even more shallower when compared to both the MS consensus and the best fit of \cite{2012ApJ...754L..29W} at similar redshift bin (1 < z < 1.5). We observe that the mean SFR for all sources do not agree with the yellow region at most M$_{*}$ bins (9.6 $\leq$ log M$_{*}$ < 10.1 and 10.1 $\leq$ log M$_{*}$ < 10.6 as well as at 11.1 $\leq$ log M$_{*}$ <  11.6).  

In Figure \ref{fig: all_sf-ms} ({\it left-hand panel}), we overall might detect larger numbers of MS galaxies with lower stellar mass (log M$_{*}$  < 10) in the first two redshift bins (0.10 $\leq$ z < 0.47 and 0.47 $\leq$ z < 0.83) where the best-fit for the MS seem to be the upper benchmark lines at z = 0.5 and z = 1, respectively. On the other hand, at higher redshift bin (0.83 $\leq$ z $\leq$ 1.20) we might probe the bulk of the MS galaxies mostly at the high mass end (log M$_{*}$  < 10) where the best-fit consensus (z=1) fits well the MS galaxies.

\subsection{Environments versus galaxy types}
In Figure \ref{fig: all_SFR}, we present the average log SFR of our (both z and M$_{*}$) binned sample where the top row represents the three different environments (field, filament, cluster) and the bottom row represents the three different galaxy types (satellite, central, isolated).
We note that in the non-scaled x-axis in Figure \ref{fig: all_SFR} (and likewise in Figure \ref{fig: sf-all-L-env-pos}) we have plotted from the less-dense to more-dense category both for the environments and for the galaxy type in aiming to trace any trend in the distribution of galaxies for the different categories as we move from the left to the right of the plot.

\subsubsection{SFR versus z and M$_{*}$}

\begin{figure*}
\begin{multicols}{2}
    \includegraphics[width=1.0\linewidth]{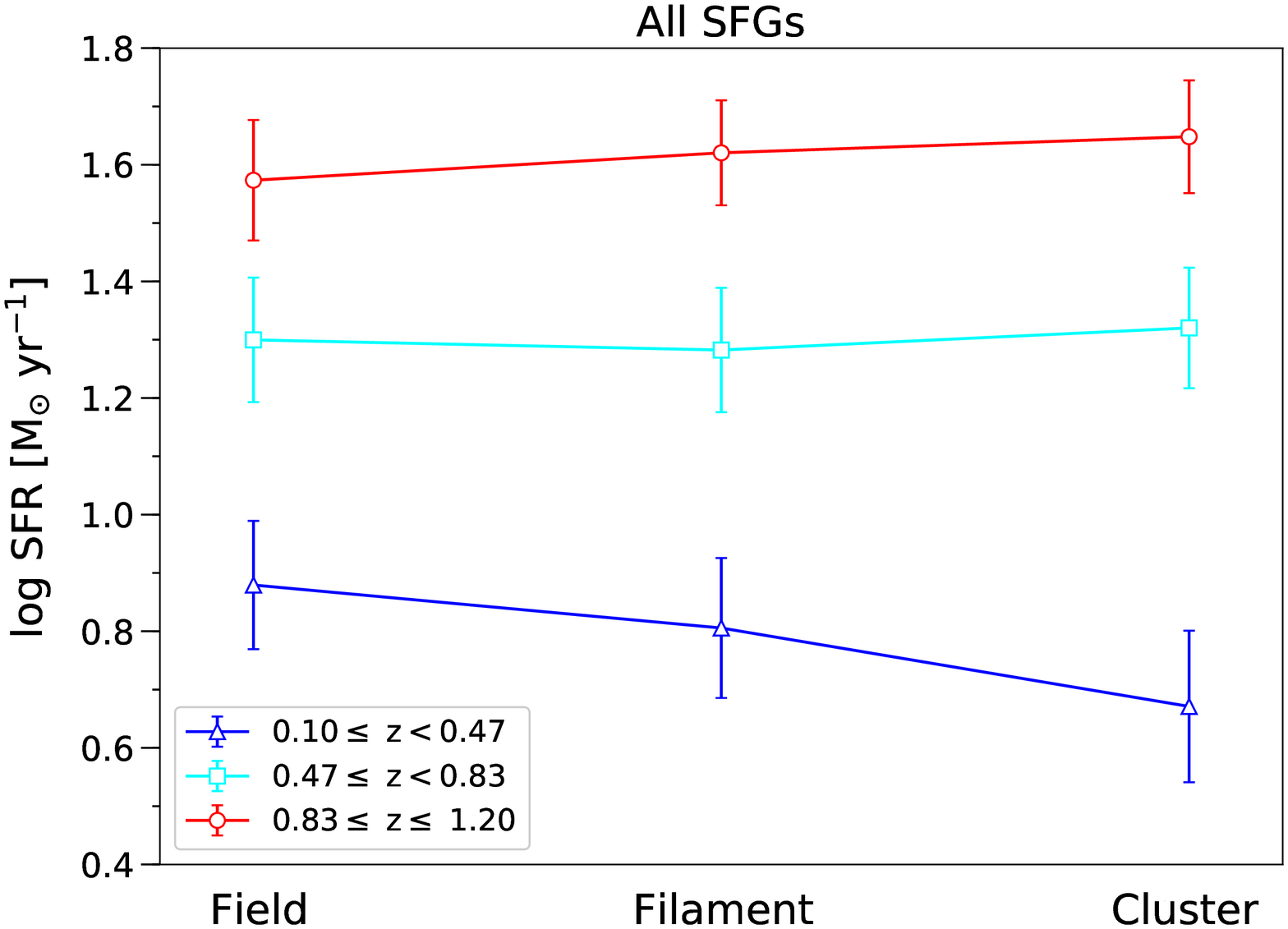}\par
    \includegraphics[width=1.0\linewidth]{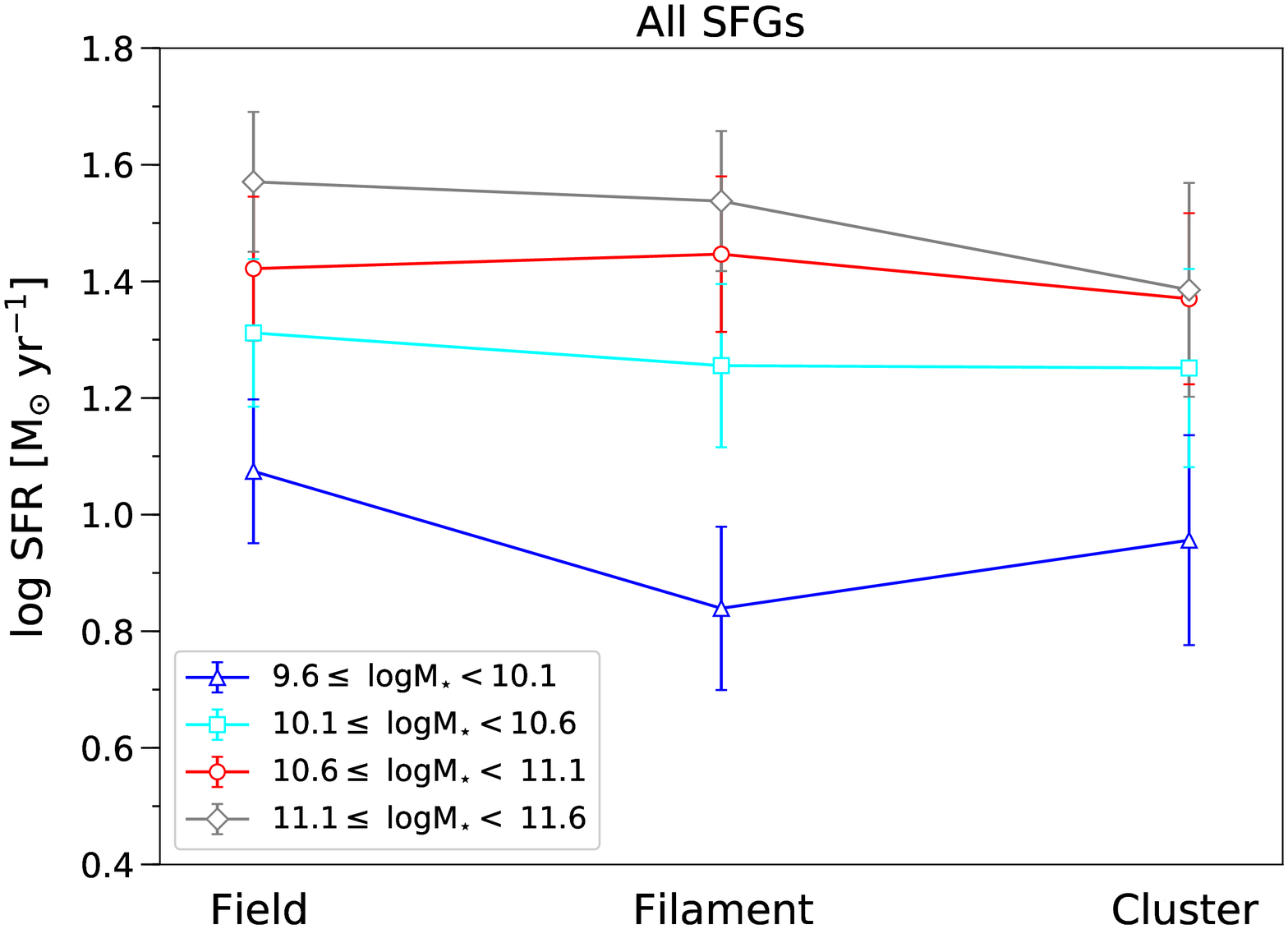}\par
\end{multicols}
\begin{multicols}{2}
    \includegraphics[width=1.0\linewidth]{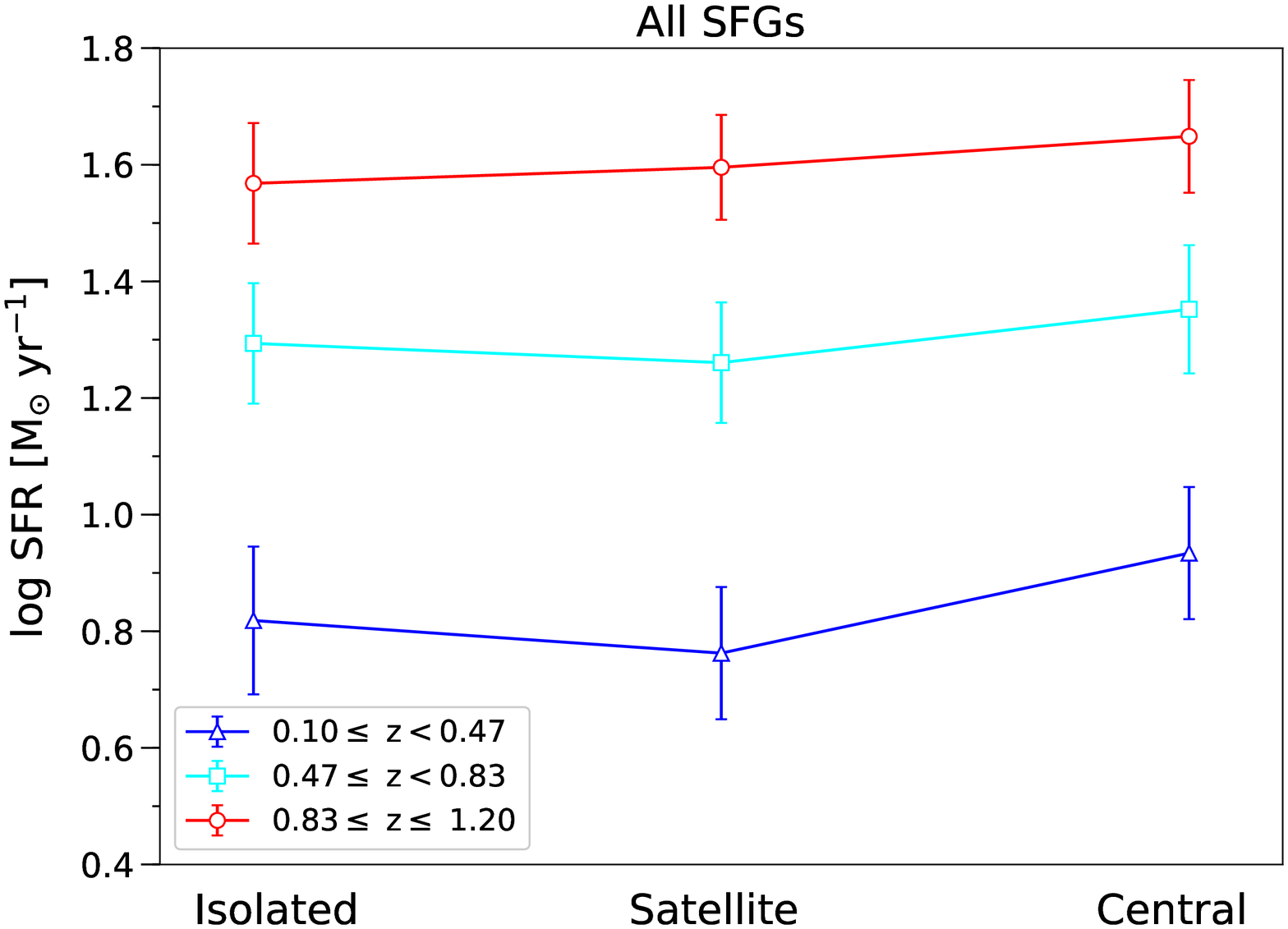}\par
    \includegraphics[width=1.0\linewidth]{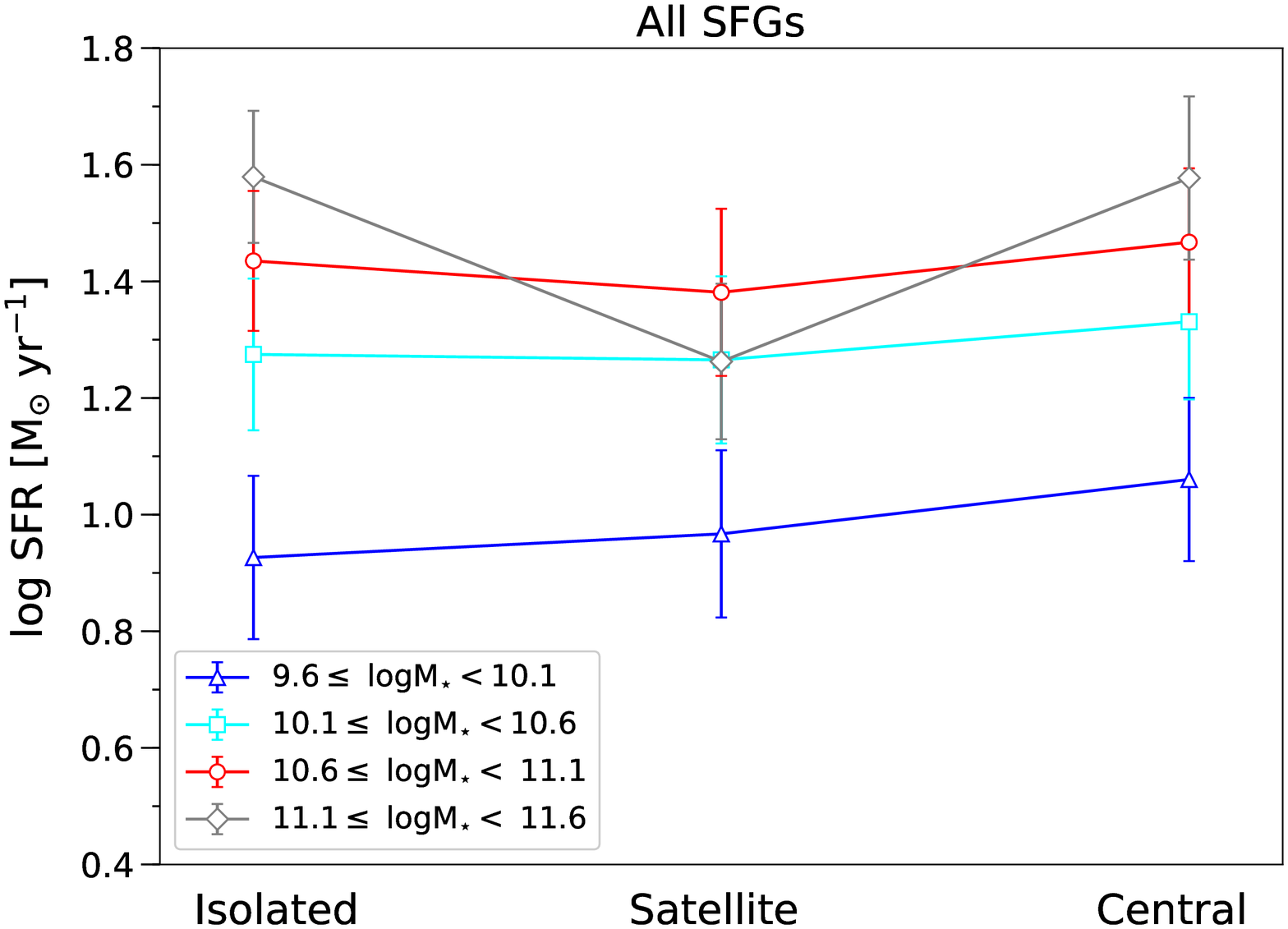}\par
\end{multicols}
\caption{{\it Top panel:} The SFR against the three different environments (field, filament, cluster) as a function of redshift bins ({\it left-hand}) and  M$_{*}$ bins ({\it right-hand}). {\it Bottom panel:} SFR against the three different galaxy types (satellite, central, isolated) as a function of redshift bins ({\it left-hand}) and  M$_{*}$ bins ({\it right-hand}). Error bars correspond to average 1$\upsigma$ errors based on the standard error of the mean. (A colour version of this figure is available in the online journal.)}
 \label{fig: all_SFR}
\end{figure*}

Figure \ref{fig: all_SFR} ({\it  top panel}) presents the average log SFR against the three different environments (field, filament, cluster) as a function of redshift  ({\it left-hand}) and  M$_{*}$  ({\it right-hand}). The bottom panel of Figure \ref{fig: all_SFR} shows the average log SFR against the three different galaxy types (isolated, satellite, central) as a function of redshift ({\it left-hand}) and  M$_{*}$ ({\it right-hand}). Overall, Figure \ref{fig: all_SFR} confirms that SFR is declining with cosmic time, consistent with the literature and the known fact that SFR peaks at z$\sim$2 \cite[see][]{2014ARA&A..52..415M,2014MNRAS.437.3516S}.

In the {(\it top left-hand)} of Figure \ref{fig: all_SFR},  at the lower redshift bin (0.10 $\leq$ z < 0.47) we tentatively observe a reduced SFR with increasing environment density i.e. from field, filament, to cluster. While at the intermediate and higher redshift bins (0.47 $\leq$ z < 0.83 and 0.83 $\leq$ z $\leq$ 1.20), we do not see any obvious trend in SFR of SFGs from field to cluster. At the {(\it top right-hand)} of Figure \ref{fig: all_SFR}, for the lowest M$_{*}$ bin (9.6 $\leq$ log M$_{*}$ < 10.1), we might see that there is a reduced SFR from field to filament then likely a higher SFR from filament to cluster. We see no trend for the higher M$_{*}$ bin (10.1 $\leq$ log M$_{*}$ < 10.6) and a tentative trend that the log SFR might be getting lower from field to cluster for the two highest M$_{*}$ bins (10.6 $\leq$ log M$_{*}$ < 11.1 and 11.1 $\leq$ log M$_{*}$ <  11.6). In the local universe, star formation is dominated by field galaxies at different environments as may be seen in Figure \ref{fig: all_SFR}  {(\it top left-hand)} while at intermediate and higher redshift filament and cluster galaxies might be more efficient in forming stars. We note that given the error bars the average SFR of SFGs might depend on the environments particularly at the lowest redshift and stellar mass bins (0.10 $\leq$ z < 0.47, 9.6 $\leq$ log M$_{*}$ < 10.1), as in the top panel of Figure \ref{fig: all_SFR} with a tentative reduced SFR from field-to-cluster, probably implying that the MS of SFGs could vary with environments \cite[e.g.][]{2016MNRAS.455.2839E}.

Figure \ref{fig: all_SFR} {(\it bottom left-hand panel)} might show evidence of an higher SFR from isolated to central at all redshift bins. At the bottom panel of Figure \ref{fig: all_SFR} {(\it right-hand panel)}, we perhaps find a trend where SFR might be getting higher from isolated to central for the three lowest M$_{*}$ bins while the highest M$_{*}$ bin likely shows lower SFR for the satellite SFGs as compared to isolated and central SFGs. We would note that given the error bars the SFR trend could particularly be more prominent at the lowest redshift and stellar mass bins (0.10 $\leq$ z < 0.47, 9.6 $\leq$ log M$_{*}$ < 10.1) where a possibility of higher SFR from isolated-to-central might be seen in the bottom panel of Figure \ref{fig: all_SFR}.

\subsubsection{P$_{\rm 1.4GHz}$ versus z}

\begin{figure}
\centering
  \begin{tabular}{@{}c@{}}
\includegraphics[width=.474\textwidth]{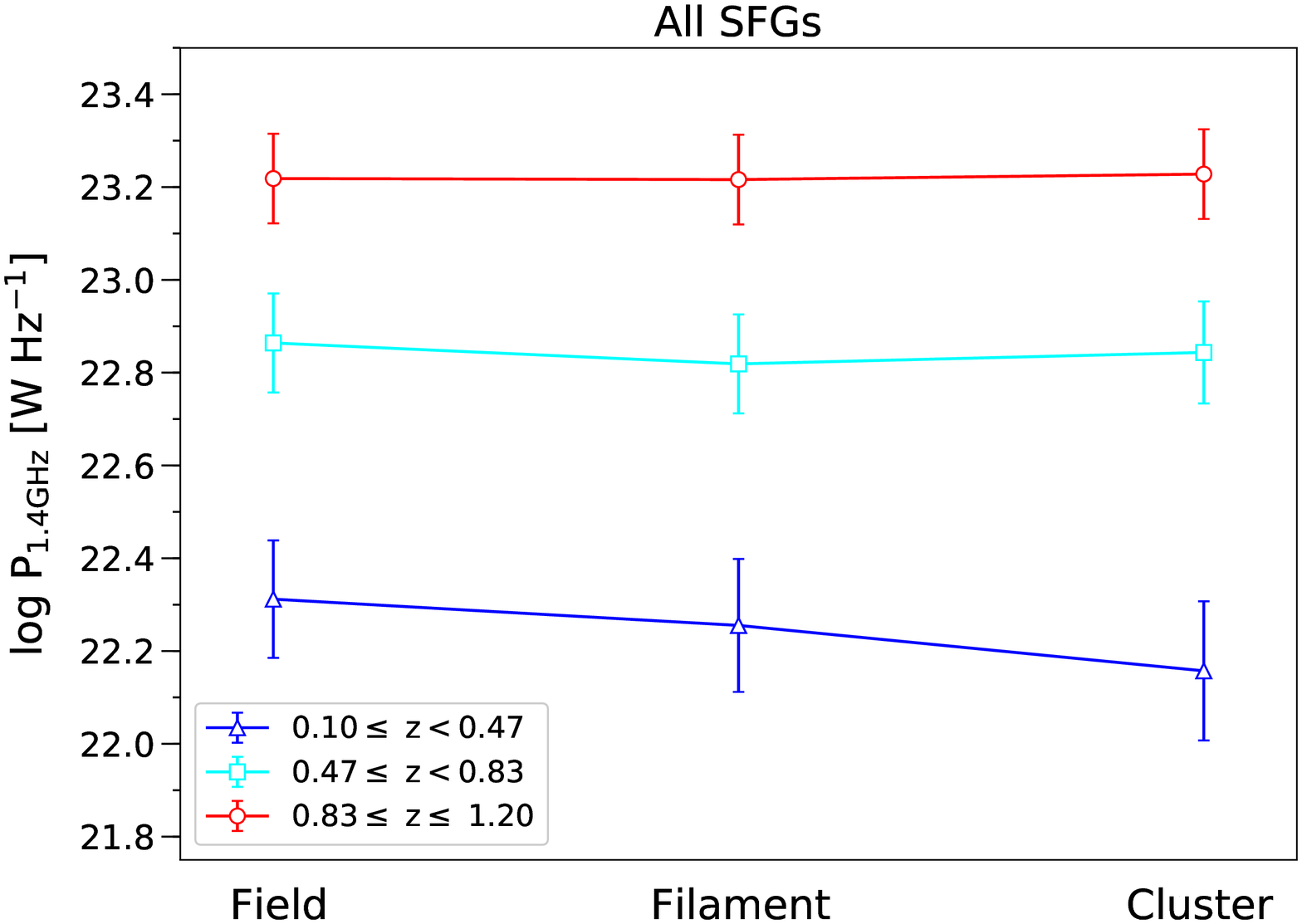} \\
\includegraphics[width=.48\textwidth]{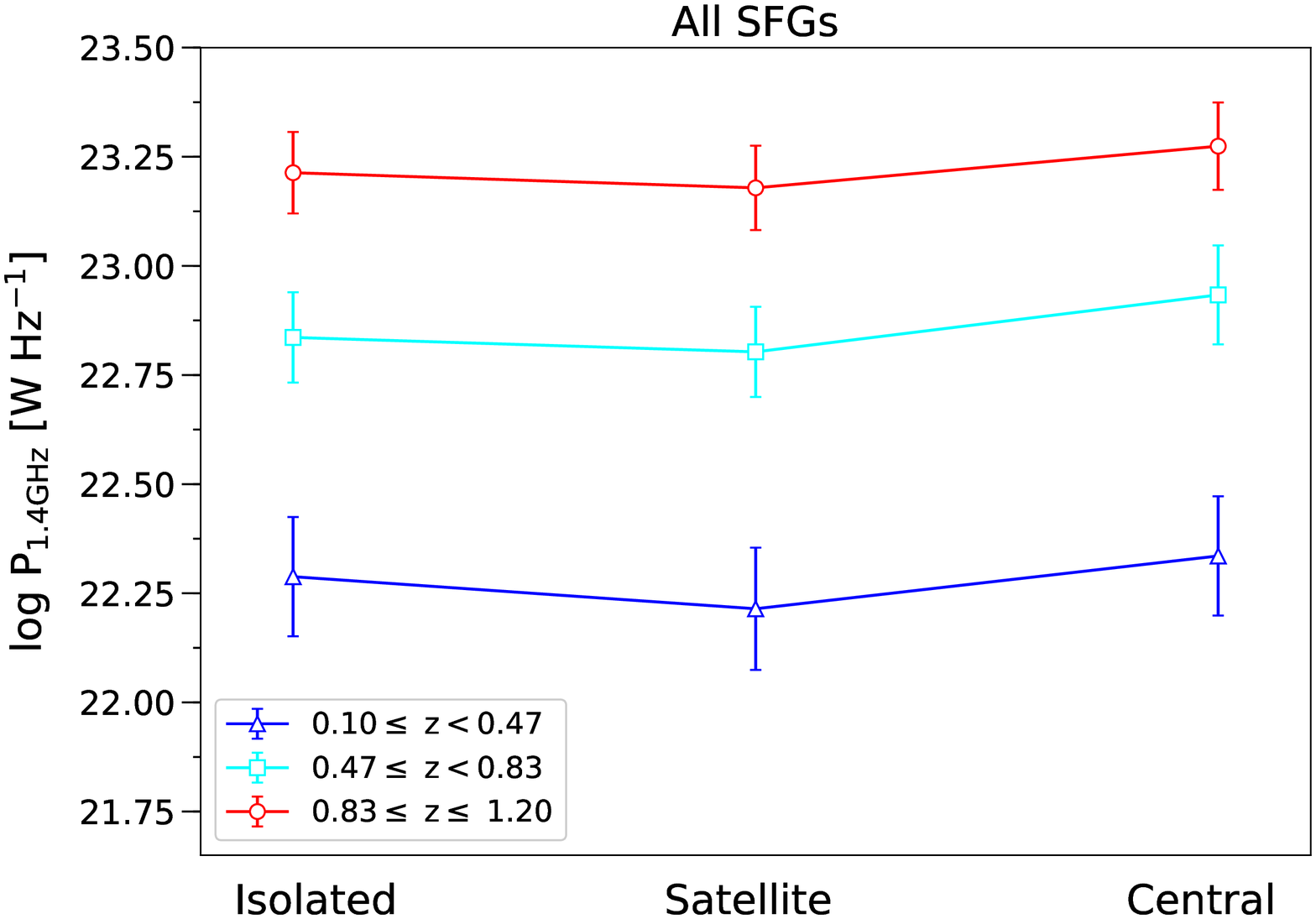} 
   \end{tabular}
 \caption{The averaged radio luminosity (P$_{\rm 1.4GHz}$) against the three different (field, filament, cluster) environments  ({\it top}) and the three different galaxy (satellite, central, isolated) types ({\it bottom}) as a function of redshift bins. Error bars correspond to average 1$\upsigma$ errors based on the standard error of the mean. (A colour version of this figure is available in the online journal.)}
 \label{fig: sf-all-L-env-pos}
 \end{figure}

In Figure \ref{fig: sf-all-L-env-pos}, we plot the average log P$_{\rm 1.4GHz}$ against the three different environments  ({\it top panel}) and the three different galaxy types ({\it bottom panel}) as a function of redshift bins. As can be seen in Figure \ref{fig: sf-all-L-env-pos}, given the errors associated with these points we effectively see no change in the log P$_{\rm 1.4GHz}$ both from field-to-cluster galaxies ({\it top panel}) and from isolated-to-central galaxies ({\it bottom panel)}.

\subsection{Possible caveats} \label{Possible caveats}
There are some possible caveats that include selection bias, sample size such as averaging a small number of sources in some bins, and completeness. 

We focus on investigation of the completeness limit of our radio selected sample by estimating the minimum SFR, i.e. SFR limits, accessible to the VLA survey. We calculated the SFR limits based on Equation 4 of \cite{2017A&A...602A...4D} using the minimum radio luminosity for each redshift bin, i.e. the SFR that correspond to the radio flux density limit.

The SFR limits are shown in the horizontal dotted lines in the left-hand panel of Figure \ref{fig: all_sf-ms} for two assumed spectral index ($\alpha$ = 0.7 and $\alpha$ = 0.8) of the SFGs populations. There is one outlier cluster galaxy at 0.10 $\leq$ z < 0.47 and one field galaxy at 0.47 $\leq$ z < 0.83 and quite a few number of the sources at 0.83 $\leq$ z $\leq$ 1.20.  For these radio sources detected under SFR limits, regardless of their environments, observed at the highest redshift bin (i.e. 0.83 $\leq$ z $\leq$ 1.20) appears to be associated with the spectral index as the two lines seem to depend on the assumed spectral index $\alpha$ as well as cosmic time. The sensitivity limits deepen as the spectral index steepen when redshift increases: possibly implying that steeper spectral index would be needed for high redshift radio sources.

We acknowledge that there might be a possibility of contamination by AGN as our sample might still have galaxies being in transition from star forming to passive with lower SFR \cite[see e.g.][]{2010ApJ...710L...1V} but still being detected as SFGs, i.e. more massive objects and at higher redshift.  We have taken advantage of the results of the multiple AGN diagnostics that excluded them already from our sample and looking at the spectra of individual galaxies may help but that is beyond the scope of the paper.

\section{Discussion} \label{Discussion} 
In Figure \ref{fig: all_sf-ms}, our comparison to the best-fit lines from the literature may indicate shallower slopes in all environments noting that the lowest and highest stellar mass bins have lower average number of sources, and the large error bars. We have measured a gradual evolution of the slope of the MS of SFGs toward shallower values at higher redshift bins (see Figure \ref{fig: all_sf-ms}). We find shallower slopes in the range of [0.2, 0.36] and normalizations in the range of [0.53, 2.90] for our radio selected sample as shown in Table \ref{tab: slopes}. Our results are in agreements with the estimated values from previous work by \cite{2009MNRAS.393..406C, 2009MNRAS.394....3D, 2010MNRAS.405.2279O, 2014MNRAS.437.3516S} who measured the slopes and normalizations  values in the range of [0.13, 0.40] and of [0.30, 3.41], respectively. The fact that we have much flatter slopes of the MS as compared to previous work of \cite{2014ApJS..214...15S} and \cite{2012ApJ...754L..29W} might be due to the result of the evolution of the completeness limits with redshift.  Other possibilities could be due to averaging small number of sources at higher mass, with relevance to that, we note that the highest M$_{*}$ bin (11.1 $\leq$ log M$_{*}$ <  11.6) we have the lowest number of sources. Other work by \cite{2011ApJ...730...61K,2012ApJ...753..114W} have also found hints of a similar trend for the slope to flatten toward z=1 or high M$_{*}$.

The variation of the slope of the MS with mass can be understood as an interplay between galaxy quenching and a depletion of the galaxy cold gas reservoir as M$_{*}$ increases \cite[e.g.][]{2017ApJ...834...39P}.  In this scenario, it is known that there is an increase with redshift of the fraction of cold molecular gas \cite[see][]{2010Natur.463..781T,2015ApJ...800...20G} and the amount of dust \cite[e.g.][]{2011MNRAS.417.1510D} available for star formation, implying that the availability of more cool gas leads to more star formation, raising the average SFR in SFGs. It has been observed that the slope of the MS becomes flatter \citep{2009MNRAS.394....3D, 2010MNRAS.405.2279O, 2014MNRAS.437.3516S} generally at higher M$_{*}$ end while the flatness seems to be a function of redshift which corroborates with our findings. 

Apart from M$_{*}$ and redshift that affect slope and the shape of the MS, there are environmental and morphological effects that play significant role at least for galaxies in the local universe \cite[e.g.][]{2019MNRAS.483.3213P}. This is found to be due to starvation of cold gas inflows in galaxies which may be truncated by hot halo as observed in the studies of local galaxies by \cite{2019MNRAS.483.3213P} which could be observed in our lower z bin.  The environmental effects such as mergers \citep{2010ApJ...721..193P,2010MNRAS.401.1613N,2011ApJ...743..159H,2012MNRAS.424.2232A,2013ApJ...778...23G,2020ApJ...889..156C} may also be observed at higher stellar mass and z which we do not see any apparent trend in Figure \ref{fig: all_sf-ms} and also the limited statistics due to lower numbers of sources in the highest mass bin does not permit to confirm this.

At fixed stellar mass, previous studies of  \cite{2008ApJ...684..888P,2014MNRAS.437..458Z} have found that environmental trends seem to weaken at higher redshift. Our results might suggest a similar weakening of environmental trends between the lowest and intermediate redshifts. As shown in Figure \ref{fig: all_sf-ms}, the MS shows the same type of flattening in all environments though the large error bars. Similarly, in Figure \ref{fig: all_SFR}, we do not see any trends as our results look flat and again given the large error bars. Overall, we point out that compared to the MSs of \cite{2012ApJ...754L..29W} and \cite{2014ApJS..214...15S}, we find the MS in all redshift bins to might be considerably flatter, irrespective of environments.

We acknowledge that the number of sources in each bin are different for both redshift and stellar mass of equally spaced of 0.26 dex for z and 0.5 dex for M$_{*}$, respectively. The three redshift bins have fairly similar number of sources (factor of $\sim$1.6 maximum difference) and numbers increase as function of redshift. However, as presented in Table \ref{tab: bins} the number of sources in the lowest and highest stellar mass bins are lower compared to the two intermediate bins (factor of $\sim$8).  By averaging a smaller number of sources in these two bins (particularly low number in cluster and at higher stellar mass) might have impact on the results in these bins when compared with respect to the other bins.

To further investigate the environmental effects, we examine how the radio luminosity of SFGs is affected by the environments and galaxy types as shown in Figure \ref{fig: sf-all-L-env-pos}. In studies of the far-infrared-radio correlation by e.g. \cite{2004ApJ...600..695R,2015MNRAS.447..168R}, it is observed that cluster galaxies have enhanced radio luminosity (relative to FIR emission) and it is attributed to be due to the impact of cluster environments on SFGs at intermediate redshift. To check this, we further investigate the environmental effects on the radio luminosity of SFGs. We did not find a clear evidence of any trends on the radio luminosity across the environments and galaxy types at all redshift bin.

\section{Conclusions} \label{Conclusions}
In this paper, we present a study of the relationship between star formation rate (SFR) and stellar mass (M$_{*}$) of star-forming galaxies, and also the relationship between environment and radio luminosity (P$_{\rm 1.4GHz}$), to shed new light on their differences with respect to environment as a function of redshift.

We use the large sample of star-forming galaxies (SFGs) from the VLA-COSMOS 3 GHz \citep{2017A&A...602A...2S} in three different environments (field, filament, cluster) for various types (isolated, satellite, central) of galaxies \citep{2015ApJ...805..121D,2017ApJ...837...16D}. We investigate for the first time the distribution of SFGs with respect to the MS consensus region from the literature via $z$ and M$_{*}$ bins, taking into account of these galaxy environments and using radio observations.  

We summarise our main results as follows:

\begin{enumerate}

\item Our results confirm that SFR is declining with cosmic time, consistent with the literature that the SFR density peaks at z$\sim$2 \cite[see][]{2014ARA&A..52..415M,2014MNRAS.437.3516S}.

\item We find that the slope of the MS for different $z$ and M$_{*}$ bins is shallower than both the MS consensus and the best-fit line of \cite{2012ApJ...754L..29W}. We measured a gradual evolution of the slope of the MS of SFGs toward shallower values at higher redshift bins, irrespective of environments, that is in agreements with previous values found by \cite{2009MNRAS.393..406C, 2009MNRAS.394....3D, 2010MNRAS.405.2279O, 2014MNRAS.437.3516S}.

\item We do not see any trend in log SFR as function of both the environments and galaxy types as given the large error bars the results could be consistent with one another. Furthermore, we thus note that overall the environment does not seem to be the cause of the flattening of MS at high stellar masses in our radio flux-limited sample.

\item We observe that log P$_{\rm 1.4GHz}$ is a function of redshift. We also do not see any trend in log P$_{\rm 1.4GHz}$ as function of both the environments and galaxy types as given the large error bars the results could be consistent with one another. As such the link between radio luminosity with galaxy types and similarly for the environments for these SFGs does not seem to depend on redshift.

\end{enumerate}

Deeper radio continuum data from the MeerKAT International GHz Tiered Extragalactic Exploration \citep[MIGHTEE,][]{2016mks..confE...6J,Taylor_2017} Survey in the COSMOS field will enable us to study larger samples of these galaxies. Furthermore, investigation of the evolution of the low mass cluster blue galaxies in the COSMOS field using the recently available data from MIGHTEE / MeerKAT complemented with ancillary data is among the aim of our upcoming manuscript. 

\section*{Acknowledgements}  
We do thank the anonymous referee for his/her valuable and constructive comments that improved the quality of this work. 
SMR wishes to thank the South African Astronomical Observatory (SAAO) for their support.
The financial assistance of the National Research Foundation (NRF) towards this research is hereby acknowledged. Opinions expressed and conclusions arrived at, are those of the author and are not necessarily to be attributed to the NRF.
This research use catalogues based on data products from observations made with ESO Telescopes at the La Silla Paranal Observatory under ESO programme ID 179.A-2005 and on data products produced by TERAPIX and the Cambridge Astronomy Survey Unit on behalf of the UltraVISTA consortium.






\bibliographystyle{mnras}
\bibliography{Randriamampandry_et_al_2020}





\bsp	
\label{lastpage}
\end{document}